\documentclass[seceq,supplement]{ptptex}

\usepackage{graphicx}

\usepackage[english]{babel} 
\usepackage{epsfig}
\usepackage{changebar}
\title{
Particle Ratios in Heavy-Ion Collisions} 

\author{A.~Tawfik}
\inst{ University of Bielefeld, P.O.~Box 100131, D-33501~Bielefeld, 
             Germany \\
       Hiroshima University, 1-7-1 Kagami-yama, Higashi-Hiroshima, Japan 
}

\abst{
In framework of statistical models, different particle
ratios at energies ranging from $3.5\;$ to $200\;$GeV are calculated. 
Assuming that the particle production takes place along the
freeze-out curve, we study the sharp peak in $K^+/\pi^+$ ratio 
observed at SPS energy. 
We study the responsibility of non-equilibrium quark
occupancy of phase space $\gamma_i$ for particle production.
Allowing $\gamma_i$ to take values other than that of
equilibrium, we got a very well description for $K^+/\pi^+$ ratio at
all energies. Using the resulting parameter set, we analyzed the 
$K^-/\pi^-$, $\Lambda/\pi^+$ and \hbox{$\Lambda/<\pi>$} ratios. We found
that the corresponding peaks all are located at
the same value of energy, $\sqrt{s_{NN}}^{(c)}\simeq7.5\;$GeV. At this
energy, the entropy per particle is singular.   
The saddle-point in entropy per particle likely refers to critical
phenomenon and change in the phase space. 
} 
\begin{document}
\maketitle


\section{\label{sec:1}Introduction}

The statistical models have been successfully used to describe the
particle production in different heavy-ion collisions. Studying 
the ratios of particle yields~\cite{Sollfrank:1999cy,Cleymans:1998yf} is of
great interest, not only to determine the freeze-out parameters, $T_{ch}$
and $\mu_B$, but also to eliminate the volume fluctuations and the dependence of the freeze-out
surface on the initial conditions. On the other hand, serious challenges rise when trying
to bring together results from different experiments, like the dependence
of $T_{ch}$ on $\mu_B$. Although each set of $T_{ch}$-$\mu_B$ originally
has been calculated by statistical models through combining various
particle ratios comparable to the experimental results, it is still
debated on the models which try to describe $T_{ch}$ vs. 
$\mu_B$~\cite{Cleymans:1999st,Braun-Munzinger:1996mq,Magas:2003wi,Tawfik:2004vv,Tawfik:2004ss,Tawfik:2005zu}. 

The $K^+/\pi^+$ ratio represents another challenge for statistical
models. The prediction of a sharp peak in $K^+/\pi^+$ ratio at SPS
energy~\cite{Gazdzicki:1998vd,Gazdzicki:2003fj} wasn't compatible with the
statistical models~\cite{Braun-Munzinger:2001as,Cleymans:1998yb}. The
latter merely expect a mild maximum, which ends up with a plateau. There
were many attempts to interpret this 
phenomenon~\cite{Letessier:2005qe,Cleymans:2005pf}.    

In this work, we study the ratios of strangeness to
non-strangeness particles. We restrict the discussion to the
kaon-to-pion and lambda-to-pion ratios. In the experimentally estimated date, we
observe that meanwhile there is a sharp peak 
in $K^+/\pi^+$ ratio, the peak of $K^-/\pi^-$ is much smooth. This
is an indication of strangeness asymmetry, which should reflect itself 
in strange hyperon production. Therefore, a sharp peak in
$\Lambda/\pi^+$ ratio is expected. The observed peak in $\Lambda/<\pi>$ is
smooth. Here $<\pi>=1.5  (\pi^++\pi^-)$. References of
experimental data can be found in~\cite{Cleymans:2005pf}

\begin{figure}[thb] 
\centerline{\includegraphics[width=10.cm]{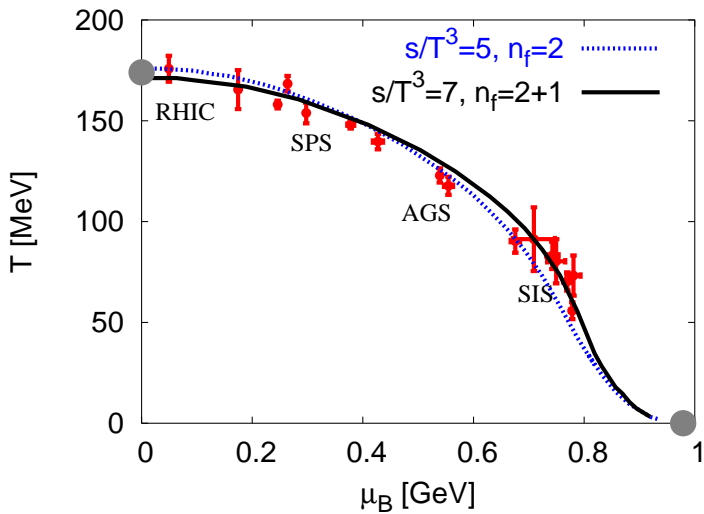}} 
\caption{\footnotesize The freeze-out parameters $T_{ch}$ and
 $\mu_B$ calculated according to the condition of constant $s/T^3$. In
 these calculations, $\gamma_q=\gamma_s=1.0$. 
\label{fig:s}
}
\end{figure}

The experimental results on $K^+/\pi^+$ ratio at different 
energies will be fitted by $\gamma_i$. $\gamma_i$ are the quark phase space
occupancy parameters. They are allowed to take values other than unity; the equilibrium
value. The subscript $i$ refers to the light and strange quark
flavors. This idea has been discussed by Johann Rafelski and his
collaborators and implemented in their statistical hadronization
model~\cite{Letessier:2005qe,Torrieri:2004zz}. Note the differences
between present work and Ref.~\cite{Letessier:2005qe,Torrieri:2004zz}! We explicitly assume that
the particle production takes place along the freeze-out curve, which is
characterized by $s/T^3=7$~\cite{Tawfik:2004vv,Tawfik:2004ss,Tawfik:2005zu}, where $s$ is
the entropy density. This condition assumes constant degrees of
freedom along the line of freeze-out, Fig.~\ref{fig:s}. Using the
parameter set from fitting $K^+/\pi^+$ vs. $\sqrt{s}$, we calculate the other  ratios. We find
that the peak in $K^+/\pi^+$ is not a unique 
phenomenon~\cite{Gazdzicki:1998vd,Gazdzicki:2003fj}. Its height and sharpness 
are obviously greater than that of the peaks observed in $K^-/\pi^-$,
$\Lambda/\pi^+$ and \hbox{$\Lambda/<\pi>$}. Another worthwhile finding
is that all peaks are located at 
almost one value of energy. This might be an indication for a critical
phenomenon. 

In the contrast to statistical models with $\gamma_i=1$, our calculation
with variable $\gamma_i$ results in an excellent agreement with the particle
ratios. For non-equilibrium freeze-out, i.e. variable $\gamma_i$, we find
that the entropy per particle $s/n$ which measures the averaged phase space
density~\cite{Gudima:1985fk,Brown:2000yf,Cramer,Tomasik:2001uz,Akkelin:2005ms}
has a maximum value located at the same collision energy as the
peaks of particle ratios do. This can be interpreted as a manifestation of
critical phenomenon. According to recent lattice QCD calculations, the
critical endpoint might be located at $\mu_B\sim0.4\;$GeV. The
corresponding $\sqrt{s}$ is much close to the energy of the particle
ratio peaks. Our
estimation for $s/n$ at RHIC energy is 
qualitatively consistent with the results given in
Ref.~\cite{Cramer,Nonaka:2005vr}. To our knowledge, there is no experimental
estimation for $s/n$ at lower energies. But there is an indication
reported in  Ref.~\cite{Cramer,Nonaka:2005vr} that $s/n$ at
SPS is higher than at RHIC. This agrees very well with our predictions, Fig.~\ref{Fig:5a}.

\section{\label{sec:2}The model}

The pressure in hadronic phase is given by the contributions of all hadron
resonances treated as a free
gas~\cite{Karsch:2003vd,Karsch:2003zq,Redlich:2004gp,Tawfik:2004sw}. At
finite temperature $T$, strangeness $\mu_S$ and iso-spin  $\mu_{I_3}$
and baryo-chemical potential $\mu_B$, the pressure of one 
hadron reads
\begin{eqnarray}
\label{eq:lnz1} 
p(T,\mu_B,\mu_S,\mu_{I_3}) &=& \frac{g}{2\pi^2}T \int_{0}^{\infty}
           k^2 dk  \ln\left[1 \pm\,
           \gamma\,
           \lambda_B \lambda_S \lambda_{I_3} e^{\frac{-\varepsilon(k)}{T}}\right],
\end{eqnarray}
where $\varepsilon(k)=(k^2+m^2)^{1/2}$ is the single-particle energy and $\pm$
stands for bosons and fermions, respectively. $g$ is the spin-isospin
degeneracy factor. $\gamma$ are the quark phase space occupancy
parameters. $\lambda=\exp(\mu/T)$ is the fugacity. $\mu$ is the chemical
potential multiplied by
the corresponding change. The particle number is given by the derivative
of Eq.~\ref{eq:lnz1} with respect to the chemical potential of interest. The
total pressure is obtained by summation all hadron resonances.

The quark chemistry is given by relating the {\it hadronic} chemical potentials and
$\gamma$ to the quark constituents. $\gamma\equiv\gamma_q^n\gamma_s^m$ with
$n$ and $m$ being the number of light and strange quarks,
respectively. $\mu_B=3\mu_q$ and $\mu_S=\mu_q-\mu_s$, with $q$ and $s$ are
the light and strange quark quantum number, respectively. The
baryo-chemical potential for the light quarks is
$\mu_q=(\mu_u+\mu_d)/2$. $\mu_S$ is calculated as a function of $T$ and
$\mu_B$ under the condition of strangeness conservation. The iso-spin
chemical potential $\mu_{I_3}=(\mu_u-\mu_d)/2$. 

In carrying out our calculations, we use full grand-canonical statistical
set of the thermodynamic parameters. Corrections due to van~der~Waals
repulsive interactions has not been taken into account~\cite{Tawfik:2004sw}. Although we do not use 
Boltzmann approximation, we can for simplicity assume it, in order show on which parameters are the
particle ratios depending. For finite iso-spin fugacity $\lambda_{I_3}$ we get 
\begin{eqnarray}
\frac{n_{K^+}}{n_{\pi^+}} \equiv \frac{K^+}{\pi^+} &\propto&
                     \lambda_s^{-1}
                     \frac{\lambda_q}{\lambda_{I_3}}\;\;\hspace*{6mm} 
                      \frac{\gamma_q}{\gamma_s}  \label{K20piplus} \\
\frac{n_{K^-}}{n_{\pi^-}} \equiv \frac{K^-}{\pi^-} &\propto&
                     \lambda_s\,
                     \left(\frac{\lambda_{I_3}}{\lambda_q}\right)\; 
                     \;\;\frac{\gamma_s}{\gamma_q}  \label{K20piminus}\\
\frac{n_{\Lambda}}{n_{\pi^+}} \equiv \frac{\Lambda}{\pi^+} &\propto&
                     \lambda_s\left(\frac{\lambda_q}{\lambda_{I_3}^2}\right)^2 
                     \hspace*{3mm}\gamma_q^2 \gamma_s
                     \label{K20pilambdapiplus}
\end{eqnarray}
The particle numbers at zero chemical potential, \hbox{$n_j(T)\simeq
  T\,m_j^2K_2(m_j/T)$}, represent the proportional factors in these 
  expressions. $n_j(T)$ is a smooth function of $T$. The fugacity
  $\lambda$ is also smooth function of $T$.  Correspondingly a 
  monotonic increase in the particle ratios is expected with  energy. 

As given in Eq.~\ref{eq:lnz1}, the statistical parameter $\gamma$ appears in the front
of Boltzmann exponential, $\exp(-\varepsilon/T)$. 
It gives the averaged occupancy of the phase space relative to
equilibrium limit. Therefore, in the equilibrium limit
$\gamma=1$. Assuming the time evolution of the system, we can describe 
$\gamma_i$ as the ratio between the change in particle number before
and after the chemical freeze-out, i.e. $\gamma_i =
n_i(t)/n_i(\infty)$. The chemical freeze-out is defined as a time scale,
at which there is no longer particle production and the collisions
become entirely elastic. 
In case of phase transition, $\gamma_i$ is expected to
be larger than one, because of the large degrees of freedom,
weak coupling and expanding phase space above the critical temperature to quark-gluon plasma.

\begin{figure}[thb] 
\centerline{\includegraphics[width=10.cm]{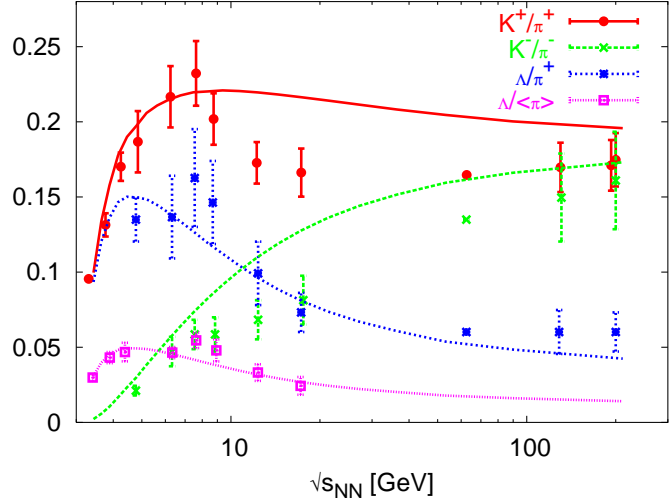}} 
\caption{\label{Fig:1a}\footnotesize
The $K^+/\pi^+$, $K^-/\pi^-$, $\Lambda/\pi^+$ and \hbox{$\Lambda/<\pi>$}
ratios at AGS ($\sqrt{s_{NN}}\leq4.84\;$GeV), SPS
($6.26\leq\sqrt{s_{NN}}\leq17.27\;$GeV) and RHIC energies
($62.4\leq\sqrt{s_{NN}}\leq200\;$GeV). Both light and strange quark occupancy parameters
$\gamma_q$ and $\gamma_s$ are assigned to unity. 
}
\end{figure}

As discussed above, thermal models using $\gamma_i=1$ can't
describe the characterized peak in $K^+/\pi^+$. We display this in
Fig.~\ref{Fig:1a}.

Allowing $\gamma_i$ to take values other than
the unity makes it possible to relate $\gamma_i$ to the collision
energy. Then we calculate $K^+/\pi^+$, Eq.~(\ref{K20piplus}). The obtained values of
$\gamma_q$ and $\gamma_s$ partly given in last two columns of Tab.~\ref{tab:1}, 
are  greater than one. This is an indication of oversaturation. To be concrete, we
emphasize that $\gamma_q/\gamma_s$ ratio first increases up to the
collision energy at which the sharp peak of $K^+/\pi^+$ ratio is
observed. Then it sharply decreases. For higher energies, it smoothly
increases. The latter indicates to the fact that the mechanism responsible for the particle
production remains unchanged from SPS to RHIC energies. In calculating the other particle ratios, we
use the same parameter set. For $K^-/\pi^-$ ratio, the ratio
$\gamma_s/\gamma_q$ multiplied by  $\lambda_s/\lambda_q$,
Eq.~(\ref{K20piminus}), is the leading parameter. For $\Lambda/\pi^+$ ratio, the occupancy
parameters $\gamma_q^2\gamma_s$, Eq.~(\ref{K20pilambdapiplus}), are
responsible for the good agreement.

\begin{table}[htp]
\begin{center}
  \begin{tabular}{|c||c||c|c||}\hline
 $\sqrt{s_{NN}}$ & $\gamma_s$ & $\gamma_q$ & $\gamma_s$
                         \\ \hline\hline 
$3.5$  & $0.8\pm0.05$ & $0.25\pm0.05$ & $0.9\pm0.02$  \\ \hline
$7.5$  & $1.0\pm0.10$ & $0.48\mp0.07$ & $1.32\pm0.12$ \\ \hline
$17$   & $0.7\pm0.07$ & $1.6\mp0.10$  & $1.43\pm0.13$ \\ \hline
$130$  & $0.8\pm0.08$ & $1.6\pm0.12$  & $1.48\pm0.11$ \\ \hline
  \end{tabular}
  \caption{\label{tab:1}\footnotesize The values of $\gamma_q$ and
  $\gamma_s$ at certain collision energies. The second column
  is related to the results depicted in
  Fig.~\ref{Fig:2a}. The corresponding $\gamma_q$ is constant and equals
 to one. The last two columns are
  the fit parameters related to Fig.~\ref{Fig:3a}.}
\end{center}
  \end{table}

\section{Results}

In Fig.~\ref{Fig:1a}, we plot the predictions from statistical models, i.e.
\hbox{$\gamma_q=\gamma_s=1$}, on the top of the experimentally estimated
results at
AGS ($\sqrt{s}\leq4.84\;$GeV), SPS ($6.26\leq\sqrt{s}\leq17.27\;$GeV) and
RHIC ($62.4\leq\sqrt{s}\leq200\;$GeV) energies. For the $K^+/\pi^+$ ratio,
there is a good agreement at AGS energy. At the SPS energy,
$\sqrt{s}\simeq10\;$GeV, there is a very mild maximum., which sets on to
a plateau. At higher
energies, this version of statistical models obviously
overestimates the $K^+/\pi^+$ ratio\footnote{For instance, at
  $\sqrt{s_{NN}}=130\;$GeV, there is a $20\%$
  overestimation. This is in contradiction with the expectations given in
  Ref.~\cite{Braun-Munzinger:2001ip}. The value of $K^+/\pi^+$ ratio
  can be indirectly calculated from the results given
  in~\cite{Braun-Munzinger:2001ip}. Accordingly, we find that
  $K^+/\pi^+\simeq0.163$ at the same energy. 
}~\cite{Cleymans:2005pf,Zschiesche:2002zr}. Even by taking into
consideration the error bars, we find that the resulting values are still 
above the ''data''.  The same behavior can be seen in 
$K^-/\pi^-$ ratio ($20\%$ overestimation at $\sqrt{s_{NN}}=130\;$GeV but
still within the error bars!). The overestimation here starts earlier
than $K^+/\pi^+$. For strange baryons, we find a decreasing population
of strange quarks with increasing energy. For $\Lambda/\pi^+$, there are underpredictions in two
energy regions. The first one is at SPS energy. 
This indicates to a peak corresponding to that of
$K^+/\pi^+$. The second one is at RHIC energy ($25\%$ underestimation at
$\sqrt{s_{NN}}=130\;$GeV, for instance). Here the resulting values lie beneath the error
bars. The same behavior is to be observed in $\Lambda/<\pi>$. 

We can so far summarize this discussion. The maxima {\it do not
  all occur at the same energy and therefore, the case for a phase transition is not very
 strong}~\cite{Cleymans:2005pf}. Furthermore, we conclude that the
particle ratios calculated by statistical models with
$\gamma_q=\gamma_s=1$, disagree
with the experimentally estimated results, especially at RHIC. \\

\begin{figure}[thb] 
\centerline{\includegraphics[width=10.cm]{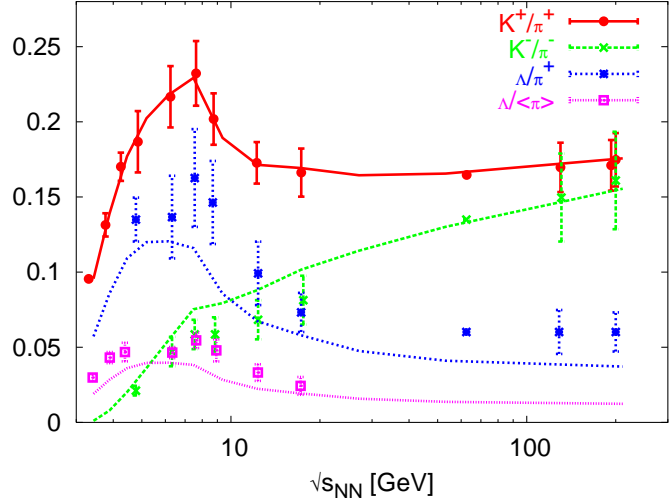}} 
\caption{\label{Fig:2a}\footnotesize
$K^+/\pi^+$, $K^-/\pi^-$, $\Lambda/\pi^+$ and \hbox{$\Lambda/<\pi>$}
 particle ratios as function of $\sqrt{s_{NN}}$. $\gamma_q$ is assigned
 to one, meanwhile $\gamma_s$ is a free parameter. Although the
 $K^+/\pi^+$ ratio has been successfully fitted, the calculation of the
 other particle ratios underestimates the experimental results. } 
\end{figure}

In Fig.~\ref{Fig:2a}, we depict the results for variable $\gamma_s$ and
constant $\gamma_q=1$. This assumes that only strangeness is out of
equilibrium. The light quark numbers are in
equilibrium. Using variable $\gamma_s$ and $\gamma_q=1$ has a long
tradition~\cite{Letessier:2005qe,Becattini:2003wp,Friese:2004av}. This
has been based on assuming strangeness saturation as a 
signature for QGP~\cite{Reflsk}. Here, we want to establish the same procedure. We
use $\gamma_s$ as a fit-parameter for $K^+/\pi^+$ 
ratio. The results are partly given in Tab.~\ref{tab:1}. We find that $\gamma_s$ always smaller than
one. This means that the strangeness is always undersaturated.  We will leave open the  
discussion on the physical interpretation, why the strangeness quantum
number has to be out of equilibrium, while the light quark quantum
numbers are explicitly in equilibrium? The results are convincing.  An
obvious underestimation for the particle ratios $\Lambda/\pi^+$ and
$\Lambda/<\pi>$ is obtained. \\ 

\begin{figure}[htb] 
\centerline{\includegraphics[width=10.cm]{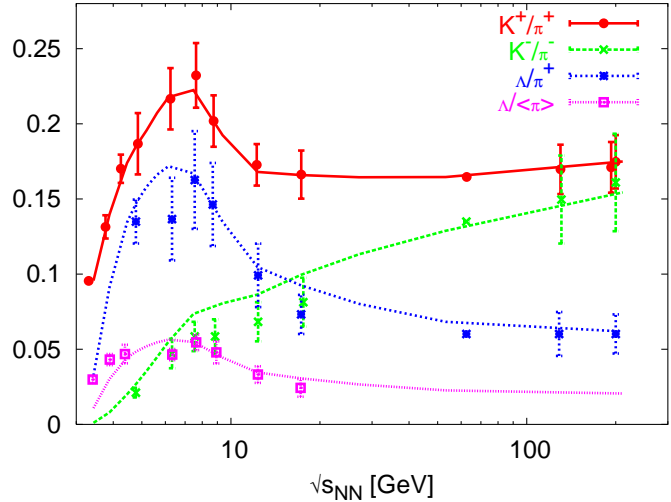}} 
\caption{\label{Fig:3a}\footnotesize
The $K^+/\pi^+$, $K^-/\pi^-$, $\Lambda/\pi^+$ and \hbox{$\Lambda/<\pi>$}
ratios at energies ranging from $3.5$ to $200\;$GeV. The experimentally estimated 
$K^+/\pi^+$ ratios have been fitted by $\gamma_q$ and $\gamma_s$. The
other particle ratios have been calculated by applying the resulting parameter set.
}
\end{figure}

In Fig.~\ref{Fig:3a}, we draw the results with varying $\gamma_i$. In
estimating the $\gamma_i$-values, we fit the experimental results on 
$K^+/\pi^+$ ratio. $\gamma_q$ is given as an input depending on
$T$~\cite{Letessier:2005qe,Rafelski:2005jc} and correspondingly on
$\sqrt{s_{NN}}$. We assumed that the particle production takes place 
along the freeze-out line, Fig.~\ref{fig:s}, at which we calculate the temperature,
$T_{ch}(\mu_B)$, according to the condition that
$s/T^3=7$~\cite{Tawfik:2004vv,Tawfik:2004ss,Tawfik:2005zu}. $s$ is 
the entropy density. All thermodynamic parameters at fixed. $\gamma_i$
are varying.  

It is interesting to note the overall agreement between our
predictions and all the experimentally estimated particle ratios. 
The parameter set obtained from fitting $K^+/\pi^+$
ratio are used to calculate the other particle ratios. There is no
fitting done here. Although 
$K^-/\pi^-$ ratio raises with the energy, we notice a weak slope within the energy region
$7<\sqrt{s_{NN}}<12\;$GeV. This might be an indication for strangeness asymmetry. 
The same behavior can be seen in the experimental data as well. While, our
calculations on this particle ratio still lie above the SPS results, the
agreement with the AGS and RHIC results is excellent. We also notice
that the $K^-/\pi^-$ ratio is not so much sensitive to the change in the light
quark occupancy parameter from $\gamma_q=1$ in Fig.~\ref{Fig:2a} to the
non-equilibrium values in this figure. 

The worthwhile finding here is that although the heights of the peaks are
different (different strangeness asymmetries), all peaks are located at almost one value of energy,
\hbox{$\sqrt{s_{NN}}^{(c)}\simeq7.5\;$GeV}.  This energy value is
corresponding to baryo-chemical potential of 
$\mu_B\simeq0.42\;$GeV~\cite{Cleymans:2005pf}. This value has been reached
by the lead beam at $40\;$AGeV at CERN-SPS. Another conclusion we can
make so far is that non-equilibrium processes in both light and strange quark
numbers are responsible for particle yields~\cite{Stoecker:2004xc}. \\

In present work, we study the responsibility of non-equilibrium quark
occupancy of phase space for particle production. We devote this
work to strangeness/non-strangeness particle ratios. 
In the following, we go one step farther in clarifying the physical reason
behind the location of the peaks. The dependence  of single-particle
entropy on the collision energy is related to the averaged phase
space density. In Boltzmann limit and for one particle
\begin{eqnarray}
\frac{s}{n} &=& \frac{\varepsilon}{T} + 1 - \frac{\mu}{T}
 \label{sOvern} 
\end{eqnarray}
where $\varepsilon$ is the single-particle energy. In this
expression, $s/n$ is related to $\varepsilon$. Apparently, $s/n$ 
becomes maximum when the chemical potential gets 
as large as the single-particle energy $\varepsilon$. As we assumed
Boltzmann limit, the maximum is unity in this case. Depending on the
chemical potential $\mu$, we can insert particles into the phase space. The
maximum occupation is reached at $\mu=\varepsilon$. This is an upper
limit. Then it becomes prohibited to insert more particles. On the other
hand, we can expect - at least theoretically - occupation values larger than
this classical upper limit, only if the phase space itself is changed. This
situation is most likely provoked by the phase transition. From this
discussion we can apparently realize that $s/n$ might play the same role
as the statistical parameter 
$\gamma$ does.

\begin{figure}[htb] 
\centerline{\includegraphics[width=10.cm]{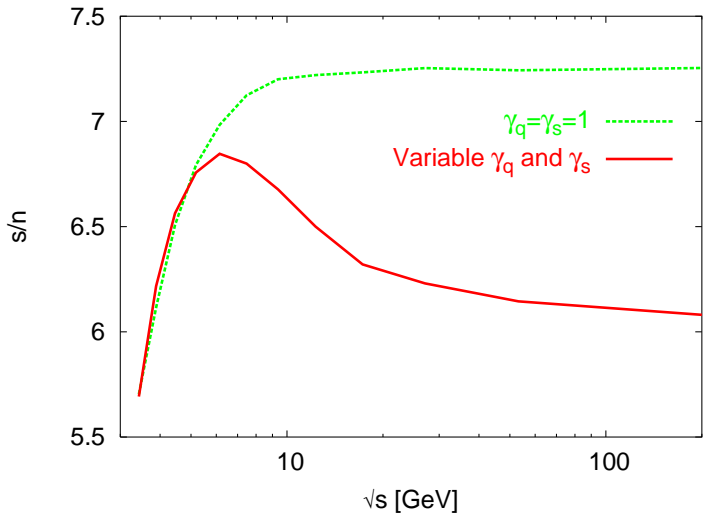}} 
\caption{\label{Fig:5a}\footnotesize
The entropy per particle $s/n$ as function of $\sqrt{s_{NN}}$. Only for
variable 
quark occupancy parameters $\gamma_q$ and $\gamma_s$ there is a singularity
in $s/n$ ratio. It is important to notice that the singularity is located at
almost the same energy as the peaks of particle ratios.  
}
\end{figure}

The results on  $s/n$ vs. $\sqrt{s_{NN}}$ are depicted in
Fig.~\ref{Fig:5a}. Again we use here many hadron resonances and full grand-canonical statistical
set of the thermodynamic parameters. In this case, the complete dependence of $s/n$ on
$T$ and $\mu$ and consequently on $\sqrt{s_{NN}}$, can straightforwardly be obtained by deriving $s$ and $n$
from Eq.~\ref{eq:lnz1}. For $\gamma_q=\gamma_s=1$, we find that
$s/n$ increases as the energy raises from AGS to low SPS
($\sqrt{s_{NN}}\leq10\;$GeV). As we saw in Fig.~\ref{Fig:1a}, a
mild maximum  in $K^+/\pi^+$ is located at the same value of
$\sqrt{s_{NN}}$. For higher energies, $s/n$ remains constant. Again,
this also is the case in Fig.~\ref{Fig:1a}. This
behavior might be an indication that a strong compensation of
the collision energy takes place in this energy region. Although we
introduce more energy in the system, the number of particles 
allowed to occupy the phase space remains constant. This is an indirect
signature that the  phase space itself remains constant.\\

For the non-equilibrium case, i.e. variable $\gamma_q$ and $\gamma_s$, we
find almost the same behavior up to $\sqrt{s_{NN}}\simeq6.5\;$GeV. At
this energy, there is a singularity in $s/n$. The
singularity might be related to a certain critical
phenomenon~\cite{Tawfik:2000mw}.
For the 
rest of SPS energies, $s/n$ rapidly decreases. Although the energy is
increased and consequently the produced particles, the single-particle
entropy decreases. This can only be understood, by assuming that the phase
space shrinks. At RHIC, $s/n$ decreases slowly with the energy. The
shrinking in phase space becomes slow. If this
model would give the correct description, we now might have for the first time a theoretical
explanation for the dependence of phase space on energy. The phase
space at SPS energy is apparently larger than the phase space at RHIC
and LHC. As discussed above, the same behaviour has been found,
experimentally~\cite{Gudima:1985fk,Brown:2000yf,Cramer,Tomasik:2001uz,Akkelin:2005ms}. 
The consequences are that the quark-gluon plasma might be produced at
SPS. And detecting its signatures at RHIC might be non-trivial.    

About the nature of critical phenomenon, we can't make any strong
statement. One might think of phenomenological models~\cite{phenom} and 
lattice QCD calculations~\cite{Fodor:2004nz,Ejiri:2003dc} for 
critical endpoint. According to lattice, the endpoint so far might be at 
$\mu_B=0.36\pm0.04\;$GeV. According to Ref.~\cite{Ejiri:2003dc},
$\mu_B\approx 0.42\;$GeV. We have to mention here, that the  
lattice calculations exclusively assume equilibrium $\gamma_q$ and
$\gamma_s$. Therefore the comparison can't be straightforward.
 

\section{\label{sec:4}Summary and discussion}

We have studied the ratios of strangeness-to-pion in dependence on the
collision energies from $3.5\;$ to $200\;$GeV. We 
assumed that the particle production takes place along the freeze-out
curve. Fixing all thermodynamic parameters and introducing quark occupancy
parameters, we have fitted $K^+/\pi^+$ ratio. Using the resulting 
parameters, we made predictions for other particle ratios. We found that the
agreement is excellent. The $\Xi/\pi$ and $\Omega/\pi$
ratios are shown in Ref.~\cite{Letessier:2005qe}. Within the statistical acceptance,
we conclude that almost all peaks are located at one value of energy.  

Different models have been suggested in order to interpret the maximum observed in
the particle ratios. Besides the statistical model at early
stage~\cite{Gazdzicki:1998vd,Gazdzicki:2003fj} there are two additional
models. The first one relates the peak to a transition from baryon-rich
to meson-rich hadronic matter~\cite{Cleymans:2005pf}. According to the
second model~\cite{Letessier:2005qe}, the peak separates a high
entropy phase from a low entropy one.   

As shown, $\gamma_q$ and $\gamma_s$ played the ace in reproducing
the particle ratios at all energies. This implies that the
particle production is due to non-equilibrium processes in both light and
strange quarks. Assuming equilibrium in only one of these quantum numbers
is not able to produce the particle ratios as given in Fig.~\ref{Fig:2a}.

We introduced $s/n$ as a thermodynamic quantity related to the
statistical parameters $\gamma$.
In Fig.~\ref{Fig:5a}, we found that the sharp peaks of the particle
ratios are associated with a singularity in the entropy per
particle. The latter can be understood by assuming a rapid modification
in the phase space. The entropy
per particle seems to play the same role as the statistical parameter $\gamma$
does. The dependence of phase space on energy is essential to access the
phase transition.


\begin{thebibliography}{99}

\bibitem{Sollfrank:1999cy}
J.~Sollfrank, U.~W. Heinz, H.~Sorge, and N.~Xu.
\newblock {\em J. Phys.}, G25:363, 1999.

\bibitem{Cleymans:1998yf}
H.~Cleymans, J.~Oeschler and K.~Redlich.
\newblock {\em J. Phys.}, G25:281--285, 1999.

\bibitem{Cleymans:1999st}
J.~Cleymans and K.~Redlich.
\newblock {\em Phys. Rev.}, C60:054908, 1999.

\bibitem{Braun-Munzinger:1996mq}
P.~Braun-Munzinger and J.~Stachel.
\newblock {\em Nucl. Phys.}, A606:320--328, 1996.

\bibitem{Magas:2003wi}
V.~Magas and H.~Satz.
\newblock {\em Eur. Phys. J.}, C32:115--119, 2003.

\bibitem{Tawfik:2004vv}
A.~Tawfik.
\newblock {\em J. Phys.}, G31:S1105--S1110, 2005.

\bibitem{Tawfik:2004ss}
A.~Tawfik.
\newblock J.~Phys.~G~{\bf 31}~S1105~(2005)

\bibitem{Tawfik:2005zu}
A.~Tawfik.
\newblock Nucl.~Phys.~A~{\bf 764}~387~(2006)

\bibitem{Gazdzicki:1998vd}
M.~Gazdzicki and M.~I. Gorenstein.
\newblock {\em Acta Phys. Polon.}, B30:2705, 1999.

\bibitem{Gazdzicki:2003fj}
M.~Gazdzicki.
\newblock {\em J. Phys.}, G30:S161--S168, 2004.

\bibitem{Braun-Munzinger:2001as}
P.~Braun-Munzinger, J.~Cleymans, H.~Oeschler, and K.~Redlich.
\newblock {\em Nucl. Phys.}, A697:902--912, 2002.

\bibitem{Cleymans:1998yb}
H.~Cleymans, J.~Oeschler and K.~Redlich.
\newblock {\em Phys. Rev.}, C59:1663, 1999.

\bibitem{Letessier:2005qe}
J.~Letessier and J.~Rafelski.
\newblock nucl-th/0504028.

\bibitem{Cleymans:2005pf}
J.~Cleymans, H.~Oeschler, K.~Redlich, and S.~Wheaton.
\newblock {\em Phys. Lett.}, B615:50--54, 2005.

\bibitem{Torrieri:2004zz}
G.~Torrieri, {\it et. al.}, 
\newblock nucl-th/0404083.

\bibitem{Gudima:1985fk}
K.~K. Gudima, V.~D. Toneev, G.~Ropke, and H.~Schulz.
\newblock {\em Phys. Rev.}, C32:1605--1611, 1985.

\bibitem{Brown:2000yf}
D.~A. Brown, S.~Y. Panitkin, and G. Bertsch.
\newblock {\em Phys. Rev.}, C62:014904, 2000.

\bibitem{Cramer}
J.~Cramer.
\newblock {\em NUCKLEONIKA}, 49:S41--S44, 2004.

\bibitem{Tomasik:2001uz}
B. Tomasik and U. ~W. Heinz.
\newblock {\em Phys. Rev.}, C65:031902, 2002.

\bibitem{Akkelin:2005ms}
S.~V. Akkelin and Yu.~M. Sinyukov.
\newblock nucl-th/0505045.

\bibitem{Nonaka:2005vr}
C.~Nonaka, B.~Muller, S.~A. Bass, and M.~Asakawa.
\newblock {\em Phys. Rev.}, C71:051901, 2005.

\bibitem{Karsch:2003vd}
F.~Karsch, K.~Redlich, and A.~Tawfik.
\newblock {\em Eur. Phys. J.}, C29:549--556, 2003.

\bibitem{Karsch:2003zq}
F.~Karsch, K.~Redlich, and A.~Tawfik.
\newblock {\em Phys. Lett.}, B571:67--74, 2003.

\bibitem{Redlich:2004gp}
K.~Redlich, F.~Karsch, and A.~Tawfik.
\newblock {\em J. Phys.}, G30:S1271--S1274, 2004.

\bibitem{Tawfik:2004sw}
A.~Tawfik.
\newblock {\em Phys. Rev.}, D71:054502, 2005.

\bibitem{Braun-Munzinger:2001ip}
P.~Braun-Munzinger, D.~Magestro, K.~Redlich, and J.~Stachel.
\newblock {\em Phys. Lett.}, B518:41--46, 2001.

\bibitem{Zschiesche:2002zr}
D.~Zschiesche, S.~Schramm, J.~Schaffner-Bielich, Horst Stoecker, and
  W.~Greiner.
\newblock {\em Phys. Lett.}, B547:7--14, 2002.

\bibitem{Becattini:2003wp}
F.~Becattini, M.~Gazdzicki, A.~Keranen, J.~Manninen, and R.~Stock.
\newblock {\em Phys. Rev.}, C69:024905, 2004.

\bibitem{Friese:2004av}
V.~Friese.
\newblock {\em J. Phys.}, G31:S911--S918, 2005.


\bibitem{Reflsk} J.~Rafelski, B.~M\"uller, Phys.~Rev.~Lett.~{\bf 48}
1066~(1982); P.~Koch, B.~M\"uller, J.~Rafelski, Phys.~Rep.~{\bf
	   142}~167~(1986);  J.~Rafelski, Phys.~Lett.~{\bf B262}~333~(1991)

\bibitem{Rafelski:2005jc}
J.~Rafelski and J.~Letessier.
\newblock hep-ph/0506140.

\bibitem{Stoecker:2004xc}
H. Stoecker, E.~L. Bratkovskaya, M.~Bleicher, S.~Soff, and X.~Zhu.
\newblock {\em J. Phys.}, G31:S929--S942, 2005.

\bibitem{Tawfik:2000mw}
A.~M. Tawfik and E.~Ganssauge.
%
A.~M. Tawfik.
\newblock hep-ph/0012022.
%
M.~I. Adamovich et~al.
\newblock {\em Heavy Ion Phys.}, 13:213--221, 2001.
%
A.~M. Tawfik.
\newblock hep-ph/0104004.
%
A.~M. Tawfik.
\newblock {\em J. Phys.}, G27:2283--2296, 2001.

\bibitem{phenom}
B.~Berdnikov and K.~Rajagopal, Phys.~Rev.~D~{\bf 61},~105017~(2000),\\
C.~Nonaka and M.~Asakawa, Phys.~Rev.~C~{\bf 71},~044904~(2005).

\bibitem{Fodor:2004nz}
Z.~Fodor and S.~D. Katz.
\newblock {\em JHEP}, 04:050, 2004.

\bibitem{Ejiri:2003dc}
S. Ejiri et~al.
\newblock {\em Prog. Theor. Phys. Suppl.}, 153:118--126, 2004.



\end{thebibliography}

\end{document}